\documentclass[
 reprint,
 amsmath,
 pre,
 amssymb,
 nofootinbib,
 aps,
 longbibliography
]{revtex4-1}

\usepackage[pdftex]{graphicx}
\usepackage{wasysym}
\usepackage{amsfonts}
\usepackage{ifsym}
\usepackage{pifont}
\usepackage{footmisc}
\usepackage[normalem]{ulem}
\usepackage{enumitem}
\usepackage{graphicx}

\makeatletter
\newlength{\apb@width}
\newcommand{\autoparbox}[2][c]{\settowidth{\apb@width}{#2}\parbox[#1]{\apb@width}{#2}}

\makeatother

\newcommand{\namedref}[2]{\hyperref[#2]{#1~\ref*{#2}}}

\def\be#1\ee{\begin{align}#1\end{align}}

\newcommand{\Csphere}{{}^\bullet\kern-1.2pt C}
\newcommand{\Ctorus}{{}^\circ\kern-1.2pt C}

\newcommand{\COMMENT}[1]{}

\newcommand{\neqa}{\nonumber\end{eqnarray}}

\newcommand{\<}{{\langle}}
\renewcommand{\>}{{\rangle}}

\newcommand{\re}{\relax{\rm I\kern-.18em R}}

\def\su2{{SU(2)}}

\def\[{\left[}
\def\]{\right]}

\def\({\left(}
\def\){\right)}
\def\[{\left[}
\def\]{\right]}

\def\<{\langle}
\def\>{\rangle}

\def\i2{\frac{i}{2}}

\def\2F1{\,_2{\rm F}_1}

\usepackage[usenames,dvipsnames]{xcolor}
\usepackage{color}
\usepackage{graphicx}
\usepackage{dcolumn}
\usepackage{bm}
\usepackage{hyperref}

\usepackage{cancel}
\usepackage{ctable}
\usepackage{booktabs}

\newcolumntype{L}[1]{>{\raggedright\let\newline\\\arraybackslash\hspace{0pt}}m{#1}}
\newcolumntype{C}[1]{>{\centering\let\newline\\\arraybackslash\hspace{0pt}}m{#1}}
\newcolumntype{R}[1]{>{\raggedleft\let\newline\\\arraybackslash\hspace{0pt}}m{#1}}

\newcommand{\beq}{\begin{equation}}
\newcommand{\eeq}{\end{equation}}
\newcommand{\beqq}{\begin{equation*}}
\newcommand{\eeqq}{\end{equation*}}
\newcommand\beqa{\begin{eqnarray}}
\newcommand\eeqa{\end{eqnarray}}
\newcommand\beqaa{\begin{eqnarray*}}
\newcommand\eeqaa{\end{eqnarray*}}
\newcommand\bea{\begin{array}}
\newcommand\eea{\end{array}}

\newcommand\myeq{\mathrel{\overset{\makebox[0pt]{\mbox{\normalfont\tiny\sffamily $d=2$}}}{=}}}

\begin{document}

\title{Nonperturbative Anomalous Thresholds}

\author{Miguel Correia}

\affiliation{CERN, Theoretical Physics Department, CH-1211 Geneva 23, Switzerland}
\affiliation{Fields and Strings Laboratory, Institute of Physics, École Polytechnique Fédérale de Lausanne, Switzerland}

\begin{abstract}
\noindent Feynman diagrams (notably the triangle diagram) involving heavy enough particles contain branch cuts on the physical sheet - anomalous thresholds - which, unlike normal thresholds and bound-state poles, do not correspond to any asymptotic $n$-particle state. ``Who ordered that?" We show that anomalous thresholds arise as a consequence of established S-matrix principles and two reasonable assumptions: unitarity below the physical region and analyticity in the mass. We find explicit nonperturbative formulas for the anomalous threshold singularity and test them against the Coleman-Thun poles of the exactly solvable $E_8$ integrable model.

\end{abstract}

\maketitle

\section{Introduction}

Despite the intricate analytic structure of scattering amplitudes most singularities have a clear on-shell explanation. A basic consequence of unitarity is the presence of normal thresholds: branch cuts on the physical sheet starting at energies where particle production occurs, i.e. where intermediate states can be made on-shell (see Fig. 1). While rigorous unitarity only holds in the physical scattering region, perturbation theory indicates that the exchange of states lighter than the physical threshold should still correspond to singularities, such as the usual simple poles (single particles going on-shell). These can be accounted for by \emph{extended} unitarity \cite{Mandelstam:1960zz, olive1963unitarity, PhysRev.135.B745, boyling1964hermitian, Eden:1966dnq, Homrich:2019cbt, Karateev:2019ymz, Guerrieri:2020kcs,Hannesdottir:2022bmo}, which assumes that unitarity remains valid \emph{below} the physical region (see Fig. 1).
\par
Yet, when going beyond the $2 \to 2$ scattering of the lightest particle (i.e. $2 \to 2$ scattering of heavier particles or generic multi-particle scattering) one typically encounters singularities in perturbation theory which are not captured by extended unitarity (as defined above). The prototypical example is the triangle diagram in Fig. 1. When the mass $M$ of the heavier particle exceeds $\sqrt{2}$ of the mass $m$ of the lightest particle (and remains stable),
\be
\label{eq:anomcond}
2 m > M > \sqrt{2}m,
\ee
the triangle singularity occurs on the physical sheet below the normal thresholds, at $s = a$, where
\be
\label{eq:a}
a = 4 M^2 - M^4/m^2 \, <  \, 4m^2.
\ee
This is known as an \emph{anomalous threshold} \cite{karplus1958spectral,karplus1959spectral,Nambu:1958zze,cutkosky1961anomalous}. 
\par
\begin{figure}[h]
    \centering
\includegraphics[scale=0.75]{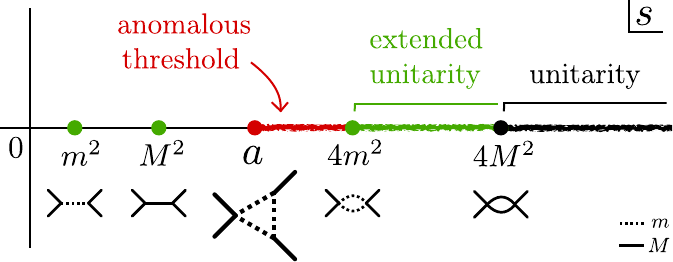}
    \caption{Complex $s$-plane for $MM \to MM$. Physical scattering occurs for $s \geq 4M^2$ where unitarity applies. Typically, unitarity is extended below to find further normal thresholds and simple poles, but not anomalous thresholds \cite{boyling1964hermitian}.}
    \label{splane}
\end{figure}
\par
Anomalous thresholds have a physical interpretation for certain composite objects.\footnote{Namely, for loose bound states, such as the deuteron, whose mass $M = 2m - \epsilon$, with binding energy $\epsilon \ll m$, satisfies the relation \eqref{eq:anomcond}. 
It follows from non-relativistic quantum mechanics that the anomalous threshold is related to the spatial extension of the bound state's wavefunction \cite{martin1961selected,barton1965introduction} (see also \cite{sasha} for a recent treatment).}
 They are nonetheless present regardless of compositeness, and they have physical consequences. In the absence of poles, the anomalous threshold is the closest singularity to the crossed channel physical region (where $s\leq 0$ is a scattering angle, see Fig. 1), and therefore controls the decay at large impact parameters of the amplitude and the corresponding Froissart bound on the asymptotic cross-section \cite{sasha}.\footnote{The triangle singularity can also affect the direct channel (even in the non-anomalous regime where it is on the second sheet). It appears to be in the origin of threshold enhancements or resonance-like effects in some processes involving exotic hadrons \cite{Guo:2014iya,Szczepaniak:2015eza,Liu:2015taa,Mikhasenko:2015vca,Guo:2016bkl,Bayar:2016ftu}. See \cite{Guo:2019twa} for a list of processes where the triangle singularity is suspected to play a role (of note the $a_1(1420)$ `peak' \cite{Mikhasenko:2015oxp,Aceti:2016yeb}).}
\par
Perturbation theory indicates that anomalous thresholds are not only present in $2 \to 2$ scattering of heavy enough particles, but also in generic multi-particle scattering (even of the lightest particle, say $ m m \to m m m $) \cite{boyling1964hermitian,cutkosky1961anomalous}. The absence of anomalous thresholds (or other exotic singularities \cite{PhysRev.115.1741,doi:10.1063/1.1703752,Eden:1966dnq}) in Feynman diagrams of $2 \to 2$ scattering of the lightest particle (here $m m \to m m$) is quite special and  motivates the hypothesis of lightest particle maximal analyticity (LPMA) \cite{Correia:2021etg}. 
\par
LPMA has important practical implications:  Bounds on physical observables such as Wilson coefficients can be determined since \emph{all} branch cuts of the amplitude are normal thresholds and are directly constrained by unitarity and positivity. On the other hand, the presence of an unconstrained branch cut, such as an anomalous threshold, would be a serious obstacle to finding bounds in general, as the amplitude can oscillate wildly across this cut (where it technically is a distribution). Given the renewed interest on the numerical S-matrix bootstrap \cite{Guerrieri:2020kcs,Karateev:2019ymz,Paulos:2016but,Doroud:2018szp,Paulos:2017fhb,He:2018uxa,Cordova:2018uop,Guerrieri:2018uew,Homrich:2019cbt,EliasMiro:2019kyf,Paulos:2018fym,Bercini:2019vme,Cordova:2019lot,Kruczenski:2020ujw,Guerrieri:2020bto,Hebbar:2020ukp,Sinha:2020win,Guerrieri:2021ivu,Tourkine:2021fqh,Karateev:2022jdb,EliasMiro:2021nul,He:2021eqn,Guerrieri:2021tak,Chowdhury:2021ynh,Chen:2022nym,Miro:2022cbk, Guerrieri:2022sod,Haring:2022sdp,Correia:2022dyp} (see \cite{Kruczenski:2022lot} for a recent overview), understanding whether anomalous thresholds can be constrained, and how, is paramount.
\par

\par
Anomalous thresholds also famously show up in integrable models, where they are typically known as Coleman-Thun poles \cite{Coleman:1978kk}. Contrarily to bound-state poles, Coleman-Thun poles are not necessarily simple.
For example, in the $E_8$ Toda field theory, the scattering of the heaviest particle has two Coleman-Thun poles of order twelve \cite{Zamolodchikov:1989fp,1989PhLB..226...73H}. Integrable models are exact nonperturbative solutions. However, to our knowledge, current understanding of Coleman-Thun poles as anomalous thresholds is only based on Feynman diagrams and corresponding Landau analysis \cite{Delfino:2003yr}.
\par
Here we take a nonperturbative approach. In section \ref{sec:triangle} we briefly review how the anomalous threshold arises in the triangle diagram. The main argument is described in section \ref{sec:main} where the anomalous threshold of the $m m \to M M$ process in $d = 2$ is studied and checked against the $E_8$ integrable model. In section \ref{sec:gen} we generalize the previous argument to $d > 2$ and, finding a match with old results on the discontinuity on the anomalous threshold of the $m m \to M M$ in $d=4$. In section \ref{sec:MM} we propose, for the first time, a nonperturbative formula for the anomalous threshold of the $M M \to M M$ process. For simplicity, we consider $d = 2$ and compute the Coleman-Thun double pole residue of the $M M \to M M$ process. We check against the $E_8$ integrable model. Finally, in section \ref{sec:conclusion} we compare with previous literature and discuss possible generalizations of our work.

\section{The triangle diagram}
\label{sec:triangle}

It is instructive to take a brief look at anomalous thresholds in perturbation theory before we proceed. The case of triangle diagram has been studied many times (see e.g. \cite{Eden:1966dnq, Hannesdottir:2022bmo,sasha}). In summary, the anomalous threshold is a singularity which is already present for $M < \sqrt{2} m$ on the second sheet of the $s$ complex plane. As $M$ is increased, it comes closer to the branch point at $s = 4m^2$, encircling it at $M = \sqrt{2} m$ and coming onto the first sheet for $M > \sqrt{2} m$, as depicted below.
\begin{figure}[h]
    \centering
\includegraphics[scale=1.1]{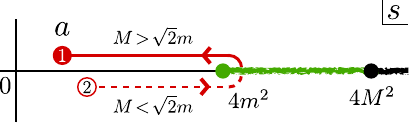}
    \label{12}
\end{figure}
\par
In $d = 2$, the anomalous threshold of the triangle diagram in fig. \ref{appA} takes the form of a simple pole. We find (see appendix \ref{app:triangle})
\be
\label{eq:triangle}
B(s \to a) = - 2 \mathcal{N}\varrho(a)  {\lambda g^2 \over s - a}, \; \text{ for } \; M > \sqrt{2} m
\ee
with 
\be
\mathcal{N} \equiv {M^4 - 2 m^2 M^2 \over m^4}, \;\;\;\;
\varrho(s) \equiv {1 \over 2 \sqrt{s} \sqrt{4m^2 - s}}.
\label{eq:Nrho}
\ee
Notice how the residue is not positive definite, given that $\lambda$ can have either sign. 
\par
\begin{figure}[h]
    \centering
\includegraphics[scale=1.35]{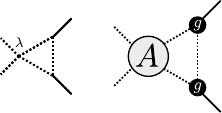}
    \caption{Left: Triangle diagram with anomalous threshold given by eq. \eqref{eq:triangle}. Right: Representation of eq. \eqref{eq:unisolB3}.}
    \label{appA}
\end{figure}
\noindent
\par
The triangle diagram is the first among infinitely many diagrams that develops an anomalous threshold at $s = a$. In particular, every diagram which can be reduced to the triangle diagram via a contraction of legs  will share the same subset of Landau singularities \cite{Eden:1966dnq,Correia:2022dyp}. All these diagrams have the following things in common with the triangle diagram:
\begin{enumerate}
    \item A two-particle cut across $2 m$.
     \item A cubic vertex $g \neq 0$ between $m m M$.
    \item Analyticity in $M^2$.
\end{enumerate}
Note that analyticity in any of the kinematic invariants - including $M^2$ - is a feature of any Feynman integral, which can be seen explicitly e.g. in the Schwinger parametrization \cite{Eden:1966dnq}. We now proceed to implement these features in a nonperturbative way. 

\section{A Nonperturbative argument}
\label{sec:main}

\par
\subsection*{Step 1: Local solution to 2-particle unitarity.} To implement the $2 m$ threshold we write down 2-particle unitarity across this branch cut. In $d = 2$ it takes a very simple form
\begin{align}
\mathrm{Disc} \, A(s) &= \rho(s) \,|A(s)|^2, \\
\mathrm{Disc} \,B(s) &= \rho(s) \, A(s) B^*(s),
\label{eq:uniAB}
\end{align}
where $\mathrm{Disc}$ is the imaginary part and $A$ and $B$ are the amplitudes for $m m \to m m$ and $mm \to MM$ scattering, respectively, and $\rho(s) \equiv \mathrm{Disc}\, \varrho(s)$ is the phase-space volume. These two equations admit a formal solution \cite{PhysRev.121.1840,Correia:2020xtr} (see appendix \ref{sec:local} for a derivation)
\begin{align}
A(s) = {\alpha(s) \over 1 -  \varrho(s) \alpha(s)  }, \;\;
B(s) = {\beta(s) \over 1 - \varrho(s) \alpha(s)}, \label{eq:unisolAB} 
\end{align}
with 
\be
\mathrm{Disc} \,\alpha(s) = \mathrm{Disc}\, \beta (s) = 0, \qquad s \in [4m^2, s_0)
\label{eq:alphabeta}
\ee
where $s_0$ is the next threshold.
\par
Solution \eqref{eq:unisolAB} effectively resums 2-particle exchanges in terms of 2mPI amplitudes $\alpha(s)$ and $\beta(s)$ (see Fig. \ref{solution} in appendix \ref{sec:local}). Importantly, it makes the analytic dependence across the $2m$ threshold manifest in terms of the 2-partice phase space volume $\varrho(s)$.
\par
\subsection*{Step 2: Inserting the cubic coupling $g$.} We can define the cubic coupling $g$ nonperturbatively as the $t$ and $u$-channel residue of the $m m \to M M$ amplitude at $m^2$,
\be
\label{eq:Bpoles}
B(s,t) \supset - {g^2 \over t - m^2} - {g^2 \over u - m^2} \;\myeq\; - \mathcal {N} {g^2 \over s - a},
\ee
where the last step is only valid in $d =2$, where $t$ and $u$ kinematically depend on $s$ (see eqs. \eqref{eq:ts2d} and \eqref{eq:us2d} in appendix \ref{app:kinematics}). Note that this is already a pole at the anomalous threshold location $s \to a$. Nonetheless, this contribution is already there for $M < \sqrt{2} m$. The anomalous threshold coming from the triangle in eq. \eqref{eq:triangle} will \emph{modify} the residue for $M > \sqrt{2} m$ in terms of $\lambda$. We wish to understand what happens for finite $\lambda$.
\par
To let solution know \eqref{eq:unisolAB} about the presence of the pole in eq. \eqref{eq:Bpoles} we can make use of the freedom in the 2PI amplitude $\beta(s)$. In particular, eq. \eqref{eq:Bpoles} implies
\be
\label{eq:beta}
\beta(s) = - {\mathcal{N} g^2 \over s - a} [1 - \alpha(a) \varrho(a)] + \beta^\text{reg}(s)
\ee
where $\beta^\text{reg}(s)$ is regular as $s \to a$. Inserting into eq. \eqref{eq:unisolAB} we have
\be
B(s) = - {\mathcal{N} g^2 \over s - a} {1 - \alpha(a) \varrho(a) \over 1 - \alpha(s) \varrho(s)} + \dots\,\,, \;\;M < \sqrt{2} m
\label{eq:unisolB2}
\ee
where `$\dots$' include regular pieces as $s \to a$.
\par
\subsection*{Step 3: Analytic continuation in $M^2$.} Eq. \eqref{eq:unisolB2} makes explicit the analytic behavior close to $s \to a$ and $s \to 4m^2$. According to eq. \eqref{eq:Nrho} it tells us that the $2m$ branch cut is of a square-root type, which is double-sheeted. To access the second-sheet one just needs to continue $\varrho(s) \to - \varrho(s)$ in eq. \eqref{eq:unisolB2}. Likewise, as $M$ increases from $M < \sqrt{2}m$ to $M > \sqrt{2}m$ we see that $a$ crosses the branch cut of $\varrho(a)$, as depicted above. Therefore, solution \eqref{eq:unisolB2} gets continued $\varrho(a) \to - \varrho(a)$ representing the anomalous threshold coming onto the first sheet. Crucially, given that $\alpha(s)$ is analytic across the $2m$ threshold according to eq. \eqref{eq:alphabeta} we assume that there is no other change to \eqref{eq:unisolB2} besides $\varrho(a) \to - \varrho(a)$.
We find that the residue at $s \to a$ gets modified to
\begin{align}
B(s\to a) &= - {\mathcal{N} g^2 \over s - a} {1 + \alpha(a) \varrho(a) \over 1 - \alpha(a) \varrho(a)}\,, \;\;M > \sqrt{2} m \notag \\
&= - {\mathcal{N} g^2 \over s - a} [1 + 2 \varrho(a) A(a)] \,, \;\;M > \sqrt{2} m
\label{eq:unisolB3}
\end{align}
where we re-expressed $\alpha(a)$ in terms of $A(a)$ using eq. \eqref{eq:unisolAB}. We see that this expression trivially reproduces the triangle diagram simple pole in eq. \eqref{eq:triangle} at leading order in perturbation theory $A(s) = \lambda + O(\lambda^2)$. 
\par
A more interesting check can be made using the $E_8$ integrable model. In integrable models, masses and couplings are fine-tuned such that particle production is absent, $B(s) = 0$. This requires the existence of another singularity at $s \to a$ with exactly the opposite residue. For example, $t-$ and $u-$ channel exchange of $M^2$ leads to a simple pole at $s \to a_{m \leftrightarrow M}$, where $m$ and $M$ are switched in $a$ given by eq. \eqref{eq:a}. Indeed, these two singularities can be made to overlap by requiring $a = a_{m \leftrightarrow M}$ which non-trivially fixes $M = {1 + \sqrt{5} \over 2} m > \sqrt{2} m$ (see figure \ref{fig:E8finetuning}). Letting $\bar{g}$ be the coupling between $m M M $ this imposes a further constraint between the residues
\be
\Big[\bar{\mathcal{N}} \bar{g}^2 + \mathcal{N} g^2 (1 + 2 \varrho(a) A(a))  \Big]_{E_8} = 0,
\label{eq:integrabletest1}
\ee
where $\bar{\mathcal{N}} = \mathcal{N}_{m \leftrightarrow M}$ in eq. \eqref{eq:Nrho}. The parameters $g$, $\bar{g}$ and $A(a)$ can be taken directly from the $E_8$ S-matrices (see appendix \ref{app:E8}), and one can verify that eq. \eqref{eq:integrabletest1} holds, providing a non-trivial test of the proposed method.

\section{Generalization to $d > 2$}
\label{sec:gen}

 Let us now describe how the previous approach gets modified for $d > 2$. In this case, the unitarity eqs. \eqref{eq:uniAB} will now retain a phase space integral \cite{Correia:2020xtr}. However, they do retain the same form in partial wave basis, which is defined in the usual way,
\be
\label{eq:pw}
A_J(s) = {1 \over 16 \pi} \int_{-1}^1 P_J(z) \, A(s,t(z)) \, dz,
\ee
and likewise for $B_J(s)$, where $P_J(z)$ is the Legendre polynomial and $z$ is the cosine of the scattering angle (see appendix \ref{app:kinematics} for the precise relation between $z$ and $t$). For concreteness, we will focus in $d = 4$, but the procedure is very similar for any $d > 2$.
\par
Step 1. of implementing the $2m$ cut therefore remains analogous with the addition of the angular momentum label $J$ in eqs. \eqref{eq:unisolAB} and rescalling the phase space factor $\varrho(s) \to (s - 4m^2) \varrho(s)$ in eq. \eqref{eq:Nrho}. The main difference is in step 2: now the poles \eqref{eq:Bpoles} in $B(s,t)$  will lead to a left-hand cut for $B_J(s)$. The partial wave integral over the poles can be easily done using the Froissart-Gribov representation (see appendix \ref{app:kinematics}). We get \eqref{eq:Bpoles}
\be
\label{eq:BpolesJ}
B_J(s) \supset {g^2 Q_J\big(Z(s)\big) \over 4 \pi  \, w(s)}, \; \text{ for even } J,
\ee
with $Z(s) \equiv (2M^2 - s) / w(s)$ and  $w(s) \equiv \sqrt{(4m^2 - s)(4M^2 - s)}$ where $Q_J(Z(s))$ is the Legendre function of the second kind which has a cut for $Z(s) \in [-1,1]$, which translates to $s \in (-\infty,a]$. We now want to impose the presence of this left cut on the solution to unitarity \eqref{eq:unisolAB}. This fixes $\beta_J(s)$ to have the form
\be
\label{eq:betaJ}
\beta_J(s) = -{g^2 \over 8 \pi}\! \int_{-\infty}^a \!\!\! {P_J\big(Z(s') \big) [1 - \alpha_J(s') \varrho(s')] \over w(s')  (s' - s)} ds' + \beta^\text{reg}_J(s)
\ee
where $\beta^\text{reg}_J(s)$ is regular across this particular cut.\footnote{Note that  $\beta^\text{reg}_J(s)$ can have other left-hand cuts. In writing \eqref{eq:betaJ} we are isolating the contribution coming from the poles \eqref{eq:Bpoles}. } It is easy to see that inserting \eqref{eq:betaJ} into the solution \eqref{eq:unisolAB} gives the correct result: $\mathrm{Disc} \,B_J(s) = \mathrm{Disc} \, B_J^\text{poles}(s)$, across the cut $s \leq a$ (making use of the relation $\mathrm{Disc}\, Q_J(z) = \pi P_J(z) / 2$).
\par
Now the continuation in the mass $M$ is more interesting. We find that as $M$ increases, $a$ encircles the $4m^2$ branch point and the integration contour \eqref{eq:betaJ} will go onto the second sheet of $\varrho(s')$. Therefore, as $a$ recedes the difference between the contours from $a \to 4m^2$ and back from $4m^2 \to a$ is nonzero, where on the return trip $\varrho(s') \to - \varrho(s')$ in \eqref{eq:betaJ}. We find that \eqref{eq:betaJ} gets corrected by
\be
\label{eq:betaJa}
\beta_J(s) \to \beta_J(s) + {g^2 \over 4 \pi} \int_a^{4m^2} {P_J\big(Z(s') \big) \alpha_J(s') \varrho(s') \over w(s') (s' - s)}
\ee
We see that this extra piece now gives a cut for $s \geq a$, i.e. an anomalous threshold. Plugging for $B_J(s)$ into \eqref{eq:unisolAB} and taking the $\mathrm{Disc}$ we get
\be
\label{eq:d4BJ}
\mathrm{Disc}\, B_J(s) = - { g^2 A_J(s) P_J\left(Z(s)\right) \over 8 \sqrt{s(4M^2 - s)} } \Theta(s - a) \, ,
\ee
Mandelstam's result \cite{Mandelstam:1960zz} is reproduced by taking $J=0$ in \eqref{eq:d4BJ}. We can invert \eqref{eq:d4BJ} back for the amplitude,
\be
\label{eq:d4B}
\mathrm{Disc}_s B(s,t(z)) = -{g^2 \int_{-1}^1 \mathcal{P}(z,z',Z(s)) \, A(s,t(z')) \, dz' \over 8 \pi \sqrt{s(4M^2 - s)}},
\ee
for $s \in [a,4m^2]$, and where $z$ is the cosine of the scattering angle and $\mathcal{P}$ is the 2-particle unitarity kernel (see appendix \ref{app:kinematics}). Eq. \eqref{eq:d4B} matches the results from \cite{boyling1966normal,goddard1969anomalous}.

\begin{figure}[h]
    \centering
\includegraphics[scale=1.30]{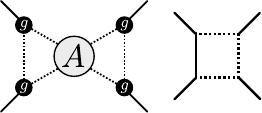}
    \caption{Left: Representation of eq. \eqref{eq:Cdp}. Right: Diagram with anomalous threshold given by eq. \eqref{eq:aboxdp} in $d = 2$.}
    \label{appA2}
\end{figure}

\section{Anomalous threshold of $M M \to M M$ in $d = 2$} 
\label{sec:MM}
Let us now see how the procedure detailed in section \ref{sec:local} applies to the anomalous threshold of $M M \to M M$ in $d = 2$, for simplicity, whose amplitude we label by $C(s)$. Repeating step 1, we write down 2-particle unitarity across $2m$
\be
\mathrm{Disc} \,C(s) = \rho(s) \,|B(s)|^2,
\ee
with solution
\be
C(s) =  \sigma(s) + { \varrho(s) \, \beta^2(s) \over 1 - \varrho(s) \alpha(s)}, \text{ with } \mathrm{Disc} \, \sigma(s) = 0
\label{eq:unisolC}
\ee
across the $2 m$ branch cut until the next threshold. Again, $\sigma(s)$ is interepreted as the 2mPI amplitude for $M M \to M M$ scattering. See appendix \ref{sec:local} for a derivation.\footnote{This solution had already been found by  Gribov \cite{Gribov:1962fx} (and partially by Oehme \cite{PhysRev.121.1840}). It also follows from the `K-matrix' solution with the second cut set to zero \cite{Pelaez:2015qba}.}
\par
Moving to step 2, we insert $\beta(s)$ given by eq. \eqref{eq:beta} into the solution \eqref{eq:unisolC}. We find
\be
C(s) = \sigma(s) + {\mathcal{N}^2 g^4 \varrho(s)[1- \alpha(a) \varrho(a)]^2 \over (s-a)^2 [1 - \alpha(s) \varrho(s) ]} + \dots
\label{eq:unisolC2}
\ee
where $\dots$ include less singular terms as $s \to a$. Now, we fix $\sigma(s)$ to cancel the double pole in $C(s)$ for $M < \sqrt{2} m$. Eq. \eqref{eq:unisolC2} then becomes
\be
C(s) = {\mathcal{N}^2 g^4 [1- \alpha(a) \varrho(a)] \over (s - a)^2} \left(\varrho(s) {1 - \alpha(a) \varrho(a) \over 1 - \alpha(s) \varrho(s)}   - \varrho(a) \right)
\ee
for $M < \sqrt{2} m$, where the factor in parenthesis vanishes as $s \to a$.
\par
Finally, the last step is to perform the continuation to $M > \sqrt{2} m$. As before, this amounts to taking $\varrho(a) \to - \varrho(a)$ in the above. We then find that the residue of $C(s)$ as $s \to a$ is no longer cancelled, but instead reads
\be
C(s \to a) =  \frac{ 2 \mathcal{N}^2 g^4  \varrho(a) [1 + 2 \varrho(a) A(a)] }{ (s-a)^2 }, 
\label{eq:Cdp}
\ee
for $M > \sqrt{2} m$, where we made use of solution \eqref{eq:unisolAB} to re-express $\alpha(a)$ in terms of $A(a)$. 
\par

A nonperturbative test can be made using the $E_8$ integrable model. However, equation \eqref{eq:Cdp} alone does not capture the double pole of the $E_8$ model (given in eq. \eqref{eq:E8dp}). There is an overlapping singularity coming the asymmetric box diagram in Fig. \ref{appA2} which becomes a double pole exactly at $M = (1 + \sqrt{5})m /2$ (see appendix \ref{app:E8}). It reads (in units of $m =1$)
\be
C_\text{abox}(s \to a) =-  {8 \sqrt{5}\,\varrho(a) + \sqrt{130+38 \sqrt{5}}\over 5} {g^2 \bar{g}^2 \over (s - a)^2}.
\label{eq:aboxdp}
\ee
If we add the two contributions from \eqref{eq:Cdp} and \eqref{eq:aboxdp}, which numerically given in eqs. \eqref{eq:npdp} and \eqref{eq:E8abox}, respectively, we find a perfect match with the double pole residue of the $E_8$ model (eq. \eqref{eq:E8dp}).

\section{Conclusion}
\label{sec:conclusion}

In this work we showed that consistency of extended unitarity and analyticity in the mass $M^2$ requires the presence of anomalous thresholds. We showcased our argument with the $ m m \to M M$ process in $d = 2$, where the anomalous threshold is found to be a simple pole given by eq. \eqref{eq:unisolB3} whose residue depends explicitly on the $m m \to m m$ nonperturbative amplitude. This formula reproduces the triangle diagram in perturbation theory and is consistent with the $E_8$ integrable model (see eq. \eqref{eq:integrabletest1}). 
\par
We then applied the method to higher dimensions ($d= 4$ was chosen for concreteness), where the anomalous threshold is a branch cut and whose discontinuity is found to be given by eqs. \eqref{eq:d4BJ}, for the partial wave, and \eqref{eq:d4B} for the amplitude. We match with Mandelstam's result for the anomalous threshold of the $J = 0$ wave \cite{Mandelstam:1960zz} whose main tool was the Muskhelishvili–Omn\`{e}s representation.\footnote{Mandelstam's argument has been reproduced and applied in different contexts \cite{blankenbecler1960anomalous,PhysRev.119.1745,PhysRev.121.1840,PhysRev.123.692,PhysRev.127.283,PhysRev.128.478,PhysRev.132.2703,PhysRev.133.B1257,gribov1963analytic,PhysRevC.13.489,Hoferichter:2013ama,Colangelo:2014dfa}.} This is a dispersion relation specific to the $ m m \to M M$ process. It assumes global analyticity and dominance of the nearest $2 m $ exchange dominates over other intermediate processes (see \cite{cutkosky1961anomalous} for a discussion). 
\par
Our method, on the other hand, only requires local analyticity assumptions and is not restricted to $ m m \to M M $ scattering. In particular, we find that the process $M M \to M M$ develops a double pole given by eq. \eqref{eq:Cdp}. A nonperturbative check of this formula was made using the $E_8$ integrable model. Because of the usual fine-tuning in the masses of integrable models required by the absence of particle production, we find other overlapping singularities. Namely, the box in fig. \ref{appA} contributes with another anomalous threshold in the $t$-channel (see appendix \ref{app:E8}) given by eq. \eqref{eq:aboxdp}. Once the two contributions are added, the $E_8$ double pole is reproduced.
\par
The nonperturbative analysis presented here corroborates Coleman and Thun's \cite{Coleman:1978kk} interpretation of the higher order poles in integrable models as anomalous thresholds. Formulas such as \eqref{eq:unisolB3} and \eqref{eq:Cdp} have been suggested before in the literature in a qualitative way with unspecified coefficients (see e.g. \cite{Delfino:2003yr, Homrich:2019cbt}). We believe this is the first time a derivation has been presented.
\par
Future directions include the analysis of the sub-leading (simple) pole behavior of $M M \to M M$ in $d = 2$ and $d > 2$, where the anomalous threshold is expected to either be of logarithm or square-root nature \cite{Landau:1959fi}. It would be interesting if the simple pole of integrable models, as in the $E_8$ model, could be reproduced via a similar set of assumptions. It would also be interesting to test the proposed formulas in non-integrable theories, such as Ising Field Theory, e.g. making use of Hamiltonian truncation (see \cite{Henning:2022xlj,Fitzpatrick:2023aqm} for recent developments).
\par

\section*{Acknowledgements}

I thank Ant\'{o}nio Antunes, Luc\'{i}a C\'{o}rdova, Hofie Hannesdottir,  Aditya Hebbar, Martin Hoferichter, Alexandre Homrich, Sebastian Mizera, Jo\~{a}o Penedones, Amit Sever, Pedro Vieira, Xiang Zhao and Alexander Zhiboedov for  useful discussions and comments on the draft. I am grateful to Alexandre Homrich and Pedro Vieira for pointing out the application to the $E_8$ integrable model. I thank the Institute for Advanced Study in Princeton and the Perimeter Institute in Waterloo for the kind hospitality while this work was being completed.  This project has received funding from the European Research Council (ERC) under the European Union’s Horizon 2020 research and innovation programme (grant agreement number 949077).

\appendix

\section{Triangle diagram in $d = 2$}
\label{app:triangle}

Here we find the anomalous threshold (Coleman-Thun pole) of the triangle diagram in fig. 2 of the main text. The Feynman parameter representation of the triangle diagram in $d = 2$ reads [5]
\be
T(s) = {\lambda g^2 \over 2\pi} \int_{0}^1 \int_0^{1-x}\!\!\!\!\! { dx \, dy  \over (s z y + M^2 x y + M^2 x z - m^2)^2}
\ee
with $z = 1 - x - y$. The answer reads
\be
\label{eq:Ts}
T(s) = -{\lambda g^2 \over m^2} {W(s) - W(a) \over s - a}
\ee
with
\be
\label{eq:W}
W(s) \equiv {2M^2 - s \over \pi \sqrt{s(4m^2 - s)}} \; \mathrm{arctan} \sqrt{s \over 4m^2 - s} 
\ee
and $a$ given by eq. 2 in the main text.\footnote{Note that $W(s) / (2M^2 - s)$ is the bubble diagram in $d=2$.}
\par
Let us look at the analyticity structure of $T(s)$. First, the function $W(s)$ only has a cut for $s \geq 4m^2$, as required by unitarity, The role of the `$\mathrm{arctan}$' is to cancel the square-root branch cut for $s \leq 0$ on the physical sheet.
\par
Now, $T(s)$ inherits this singularity structure from $W(s)$ via eq. \eqref{eq:Ts}. Note that $T(s)$ is regular at $s \to a$ because the numerator cancels the pole. In other words, no anomalous threshold exists for $M < \sqrt{2} m$.
\par
The situation changes when going to $M > \sqrt{2} m$. In this case, $a$ will go around the branch cut of the function $W(a)$ (precisely as depicted in the main text). So to write $T(s)$ explicitly in this regime we continue $W(a)$ to the second sheet, i.e. find its monodromy (to the unfamiliar reader, see dispersive argument below), which is given by 
\be
W(a) \to W(a) - {2M^2 - a \over \sqrt{a(4m^2 - a)}}.
\ee
\par
Making use of $\varrho(a) = 1/\sqrt{a(4m^2-a)}$ we have
\be
\label{eq:Tss}
T(s) \to T(s) - 2 \varrho(a) {\lambda g^2 \over m^2} {2M^2 - a \over s - a}
\ee
And we see the appearance of a pole at $s \to a$ on the physical sheet, i.e. an anomalous threshold.
\par
\begin{figure}[h]
    \centering
\includegraphics[scale=1.2]{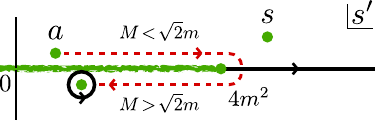}
    \caption{$s'$ complex plane. In black: integration contour. In green: singularities of the integrand. Note in particular the presence of a branch cut of $\rho(s')$ for $s' \leq 4m^2$. In red: trajectory of $a$ as $M$ is increased. At $M = \sqrt{2}$, $a = 4m^2$ pinches
 the integration contour and forces a deformation.}
    \label{deformation}
\end{figure}

Following Mandelstam [1] we can reproduce the same result directly from a dispersive representation. Taking a discontinuity of \eqref{eq:Ts}, or using unitarity directly, we get
\be
\mathrm{Disc} \, T(s) = {\lambda g^2 \over m^2} \,{s - 2M^2 \over s - a} {\Theta(s-4m^2) \over 2 \sqrt{s(s-4m^2)} }
\ee
Identifying the last factor as the phase space volume $\rho(s)$, a dispersion relation for $T(s)$ reads
\be
T(s) = {\lambda g^2 \over \pi m ^2} \int_{4m^2}^\infty {\rho(s') \over s' - s} {s' - 2M^2 \over s' - a} ds'.
\ee
It can be checked that performing the integral gives back eq. \eqref{eq:Ts}.
Now, from this representation it is clear that $T(s\to a)$ is regular, the pole $s' \to a$ in the integrand is outside the integration domain $s' \geq 4m^2$. However, if we increase $M$ we see that $a$ approaches the integration contour and forces its deformation for $M > \sqrt{2} m$. So the integral will pick up an extra term that wraps around $a$,
\be
\label{eq:contour}
T(s) \to T(s) + {\lambda g^2 \over m^2}\, {1 \over \pi  } \oint_a {\rho(s' - i \epsilon) \over s' - s} {s' - 2M^2 \over s' - a} ds'.
\ee
If $a$ is chosen to approach the contour from above, i.e. $M^2$ is given a small positive imaginary part, the contour in \eqref{eq:contour} is counter-clockwise (see fig. \ref{deformation}). Note, however that $\rho(s') \propto 1/\sqrt{s' - 4m^2}$ has a cut for $s' \leq 4m^2$, and from fig. \ref{deformation} it is clear that  $a$ will be below this cut after the continuation. So we have $\rho(s - i \epsilon) =  i \varrho(s) $ and we recover eq. \eqref{eq:Tss}.\footnote{Note that the direction in which $a$ wraps around the contour does not matter. Say it comes from below, then the contour would be clockwise but $a$ would end up on top of the branch cut of $\rho(s')$ instead and the sign difference would cancel out.} 

\begin{figure*}[t]
    \centering
\includegraphics[scale=1.27]{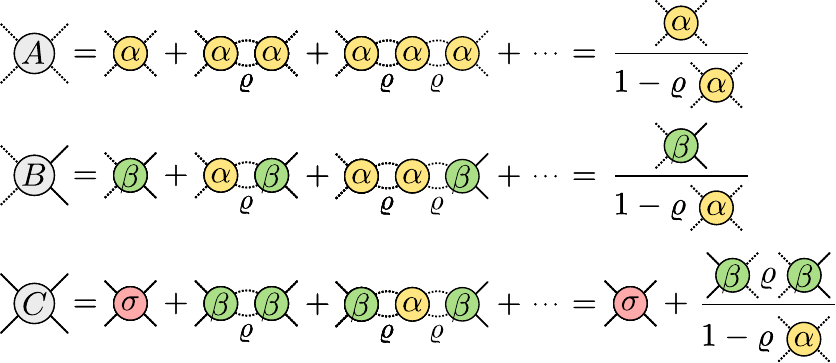}
    \caption{Graphical representation of solution to unitarity across the $2m$ cut, eqs. (7) and (21) in the main text. The sub-graphs $\alpha$, $\beta$ and $\sigma$ are $2m$-irreducible, i.e. do not contain any internal exchange of $2m$ (in the $s$-channel) meaning that $\mathrm{Disc} \,\alpha = \mathrm{Disc} \,\beta = \mathrm{Disc}\, \sigma = 0$ across this cut. }
    \label{solution}
\end{figure*}

\section{Local solution to $2 m$ unitarity}
\label{sec:local}

We will solve unitarity across the $2m$ branch cut in sequence. First for $A(s)$, then for $B(s)$, and finally for $C(s)$. Starting with $A(s)$:
\be
\mathrm{Disc} A &= \rho |A|^2 \Leftrightarrow \mathrm{Disc}  \left( {1 \over A} + \varrho \right) = 0 
\ee
meaning
\be
{1 \over A} + \varrho = {1 \over \alpha}, \text{ with } \mathrm{Disc} \,  \alpha = 0
\ee
except for possible poles of $\alpha$. This is the well-known `inverse amplitude' trick.
\par
Now, moving onto $B(s)$, we use the notation $B\pm = B(s \pm i \epsilon)$,
\be
&\mathrm{Disc}  B =  \rho B A^* \Leftrightarrow {B_+ - B_- \over 2i} = {\rho B_+ \alpha \over 1 - \alpha \varrho_-} \Leftrightarrow \\
&B_+ \left[1 - \alpha(\varrho_- + 2 i \rho)\right] - B_- \left[1 - \alpha \varrho_-\right] = 0 \Leftrightarrow \\
&B_+ \left[1 - \alpha \varrho_+ \right] - B_- \left[1 - \alpha \varrho_-\right] = 0 \Leftrightarrow \\
&\mathrm{Disc} \left[B(1 - \alpha \varrho) \right] = 0 \Leftrightarrow B(1 - \alpha \varrho) = \beta
\ee
with $\mathrm{Disc} \, \beta = 0$.
\par
Finally, we go to $C(s)$:
\be
&\mathrm{Disc}\,  C = \rho |B|^2 \Leftrightarrow \mathrm{Disc}\,  C = {\rho  \beta^2 \over |1 - \alpha \varrho|^2} \\
&\Leftrightarrow \mathrm{Disc}\, C = \mathrm{Disc}\,  \left[ {\varrho \beta^2 \over 1 - \alpha \varrho} \right]  \\
&\Leftrightarrow  C = \sigma +   \left[ {\varrho \beta^2 \over 1 - \alpha \varrho} \right], \text{ with } \mathrm{Disc}\,\sigma = 0,
\ee
where we made use of
\be
&\mathrm{Disc}\, \left[ {\varrho \over 1 - \alpha \varrho} \right] = {1 \over 2i} \left[ {\varrho_+ \over 1 - \alpha \varrho_+} - {\varrho_- \over 1 - \alpha \varrho_-} \right] = \\
&= {\varrho_+(1 - \alpha \varrho_-) - \varrho_-(1 - \alpha \varrho_+) \over 2i (1 - \alpha \varrho_+)(1 - \alpha \varrho_-)}  \\
&= {(\varrho_+ - \varrho_-)/2i \over (1 - \alpha \varrho_+)(1 - \alpha \varrho_-)} = {\rho \over |1 - \alpha \varrho|^2}.
\ee

\section{The $E_8$ integrable model}
\label{app:E8}

Here we look in detail to the Coleman-Thun/anomalous thresholds of the $E_8$ integrable model. In units where $m = 1$, the six lightest particles (out of eight) in the $E_8$ model are [59]
\be
\label{eq:E8masses}
m &= m_1 = 1, \qquad M = m_2 = 2 \cos{\pi \over 5} = {1 + \sqrt{5} \over 2} \approx 1.618, \notag \\
m_3 &= 2 \cos{\pi \over 30} \approx 1.989, \qquad m_4 = 2 m_2 \cos{7 \pi \over 30} \approx 2.405, \notag \\
m_5 &= 2 m_2 \cos{2 \pi \over 15} \approx 2.956, \qquad m_6 = 2 m_2 \cos{ \pi \over 30} \approx 3.218.
\ee
To describe the S-matrices in integrable models it useful to use the basic CDD building block which solves elastic unitarity and crossing symmetry in $d = 2$. Let us define it in an abbreviated way as
\be
\big[\mu^2\big] \equiv {\varrho(\mu^2) + \varrho(s) \over  \varrho(\mu^2) - \varrho(s) }
\ee
where $\varrho(s)$ is the phase space factor.\footnote{For the scattering of particles of different mass we have
\be
mm \to mm: \qquad \varrho(s) &= {1 \over 2 \sqrt{4m^2 - s} \sqrt{s}}, \notag \\
mm \to MM: \qquad \varrho(s) &= {1 \over 2 \sqrt{(m+M)^2 - s} \sqrt{s - (m - M)^2}}, \notag \\ 
MM \to MM: \qquad \varrho(s) &= {1 \over 2 \sqrt{4M^2 - s} \sqrt{s}}. \notag
\ee}
In the sector of the two lightest particles $m$ and $M$ we have
\be
\label{eq:SE8}
S_{mm \to mm}(s) &= \big[m^2 \big] \; \big[M^2 \big] \; \big[m_3^2 \big], \notag \\
S_{m M \to m M}(s) &= \big[m^2\big] \; \big[M^2\big] \; \big[m_3^2 \big] \; \big[m_4^2\big], \notag \\
S_{M M \to m m}(s) &= 1, \notag \\
S_{M M \to M M}(s) &= \big[m^2\big] \; \big[M^2\big] \; \big[m_4^2\big] \; \big[m_5^2\big] \; \big[m_6^2\big] \; \big[a\big]^2.
\ee
This means in particular that $m m \to m m$ has three simple poles at $m^2$, $M^2$ and $m_3^2$ (and their crossing symmetric images $s \leftrightarrow 4m^2 - s$), and likewise that $M M \to M M$ has simple poles at $m^2$, $M^2$, $m_4^2$, $m_5^2$, $m_6^2$ and a double pole at 
\be
a = 4M^2 - M^4/m^2, 
\ee
which is the anomalous threshold. Plugging the numbers we get $\sqrt{a} \approx 1.902$ so the anomalous threshold sits between the second and third particles $M  < \sqrt{a} < m_3$.
\par

\subsection*{Cancellation of the Coleman-Thun pole in $ MM \to mm$}

How is it possible that $T_{MM \to mm}(s) = 0$ if $T_{m M \to m M}(s) \neq 0$? These two objects are related by crossing symmetry in $d > 2$ but in $d = 2$ since $t = t(s)$ and $u = u(s)$ this is not necessarily the case. Let us see explicitly how $T_{MM \to mm}(s) = 0$ at ``tree level". By ``tree level" we mean just looking at how the simple poles cancel each other. 
\par
\begin{figure}[h]
    \centering
\includegraphics[scale=0.9]{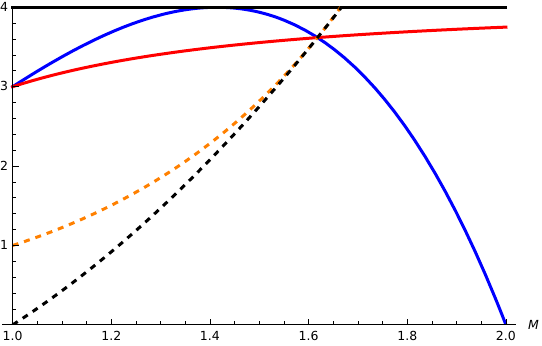}
    \caption{In black: the normal thresholds at $s = 4m^2$ and $t = 4M^2 - s = (m + M)^2$. In thick line: $s$-channel singularities in the $mm \to MM$ process. In dashed line: $t$-channel singularities in the $MM \to MM$ process. Red and blue are given by $s=a$ and $s =b$ in eq. \eqref{eq:E8poles}. Orange is given by $4M^2 - t_+$ with $t_+$ in eq. \eqref{eq:tminus}. As $M$ increases  $a = 4M^2 - M^4$ touches the $s = 4$ branch point at $M = \sqrt{2} \approx 1.41$ and comes back. When $M = {1 + \sqrt{5} \over 2} \approx 1.62$, $b$ overlaps with $a$ and cancels production of $MM \to mm$. For the $MM \to MM$ the other singularities from the $t$-channel (in dashed) and contribute to the anomalous threshold at $s \to a$. In total the 4 singularities play a role in the Coleman-Thun double pole of the $E_8$ model.}
    \label{fig:E8finetuning}
\end{figure}

The $M M \to m m$ process should be able to exchange the same particles as the `crossing symmetric' process $m M \to m M$, namely $m$, $M$, $m_3$ and $m_4$. Since $t = t(s)$ and $u = u(s)$ according to eq. \eqref{eq:ts2d} all these exchanges will give rise to poles in $s$. From eq. \eqref{eq:ts2d} we have that a pole in $t$-channel or $u$-channels at $\mu^2$ gives rise to the following pole in $s$,
\begin{align}
&t(s) = \mu^2 \;\;\text{ or }\;\; u(s) = \mu^2 \notag \\
&\implies s(\mu^2) =- {(m^2 - M^2)^2 \over \mu^2} + 2(m^2 + M^2) - \mu^2
\end{align}
For the particular values where $t(s) =\mu = m, M$ we find 
\be
\label{eq:E8poles}
&\mu = m \implies s = a =  4M^2 - {M^4 \over m^2}, \notag \\
&\mu = M \implies s = b \equiv a_{m\leftrightarrow M} = 4m^2 - {m^4 \over M^2}, 
\ee
The first value is the already studied t-channel and u-channel pole in eq. (9) of the main text. As we can see in figure \ref{fig:E8finetuning}, it wraps around the $s = 4m^2$ branch cut as $M$ increases and its residue changes as a result, i.e. becomes ``anomalous'' for $M > \sqrt{2} m$ as argued in the main text. The second value, on the other hand, does not wrap around $s = 4m^2$ as $M$ increases. 
\par
Both poles will coincide if $a = b$,
\be
\label{eq:E8m2}
4M^2 - {M^4 \over m^2} = 4m^2 - {m^4 \over M^2} \implies M = \left({1 + \sqrt{5} \over 2}\right) m,
\ee
which is precisely the mass of the second bound state in the $E_8$ model.\footnote{There are other solutions to eq. \eqref{eq:E8m2}. However, the solution in eq. \eqref{eq:E8m2} is the only one for which $M > m$. }

So, in order to have $T_{mm \to MM} = 0$, the residues coming from these two poles must cancel out. The anomalous residue was computed in the main text in  eq. (12) where, in the notation of this section, reads
\be
T_{m m \to MM}(s \sim a) = - \mathcal{N}_m {g_{mmM}^2 S_{mm \to mm}(a) \over s - a}
\label{eq:pole1}
\ee
with the relation between the amplitude and the S-matrix being, as usual, $S(s) = 1 + 2 \varrho(s) T(s)$, and 
\be
\mathcal{N}_m = {2 M^2 - a \over m^2} = \left({M \over m}\right)^4 - 2 \left({M \over m}\right)^2.
\ee
We now want to compare this with the residue coming from the other pole,
\be
- {g_{m M M}^2 \over t(s) - M^2} - {g_{m M M}^2 \over u(s) - M^2} = - {g_{m M M}^2 (s - 2m^2) \over M^2(s - b)}
\ee
so when $s \sim b$ we have
\be
\label{eq:simb}
T_{m m \to MM}(s \sim b) = - \mathcal{N}_M {g_{mMM}^2 \over s - b}, 
\ee
with
\be
\mathcal{N}_M = \left({m \over M} \right)^4 - 2\left({m \over M} \right)^2 
\ee
For these two to cancel we must have
\be
\label{eq:cancellation}
 \mathcal{N}_m \; g_{m m M}^2 \; S_{mm \to mm}(a) + \mathcal{N}_M \; g_{m M M}^2 = 0.
\ee
We can now verify that this is indeed the case in the $E_8$ model. We can extract $g_{m m M}$ by looking at the $s$-channel exchange of $M$ of the $m m \to m m$ process, i.e. 
\be
T_{m m \to m m}(s \sim M^2) = - {g_{m m M}^2 \over s - M^2}
\ee
which, making use of the S-matrices in eq. \eqref{eq:SE8}, gives
\begin{align}
\label{eq:gmmM}
&g_{m m M}^2 = \notag \\
&\sqrt{30} \sqrt{4605 + 2047 \sqrt{5}+10 \sqrt{30}\sqrt{14045 + 6281 \sqrt{5}}}
\end{align}
taking the values in \eqref{eq:E8masses}.
\par
To extract $g_{m M M}$ we look at
\be
T_{M M \to M M}(s \sim m^2) = - {g_{m M M}^2 \over s - m^2}
\ee
and find
\begin{widetext}
\be
g_{m M M}^2 &= 2 \sqrt{15} \sqrt{33814455+15122284 \sqrt{5} + 10 \sqrt{15} \sqrt{1524538572625+681794376542 \sqrt{5}}}.
\label{eq:gmMM}
\ee
\end{widetext}
\par
Likewise, we find
\be
S_{m m \to mm}(a) = 13 + 3 \sqrt{5} + 5 \sqrt{15 + 6 \sqrt{5}} 
\ee
and, finally,
\be
\mathcal{N}_m ={1 + \sqrt{5} \over 2}, \qquad \mathcal{N}_M = {1 - \sqrt{5} \over 2}.
\ee
We then verify that eq. \eqref{eq:cancellation}, or eq. (13) in the main text, is exactly satisfied. 
\par
We can also verify cancellation of the other singularities. No other anomalous threshold should exist. Any singularity coming from $t(s) = \mu^2$ or $u(s) = \mu^2$ only wraps around the $s = 4m^2$ branch cut if $M^2 > \mu^2 + m^2$ and none of the other masses $M$, $m_3$ and $m_4$ satisfy this except $m$ which gives the only anomalous threshold at $s = a$.\footnote{In perturbation theory, the condition $M^2 > \mu^2 + m^2$ for the existence of the anomalous threshold follows from requiring that the solution to the Landau equations of the triangle diagram is '$\alpha$-positive'.} 
\par
The presence of $m_3$ and $m_4$ is nonetheless necessary to cancel out the remaining singularities. In particular, we find 
\be
t(s) = m_3^2 \implies s = M^2, \qquad t(s) = m_4^2 \implies s = m^2.
\ee
It is a straightforward exercise to see that the residues also cancel each other.
\par
Finally, the singularities at $s = m_3^2$ and $s = m_4^2$ whose residues are respectively $g_{MM m_3} g_{m m m_3}$ and $g_{M M m_4} g_{m m m_4}$ are not present since $g_{M M m_3} = g_{m m m_4} = 0$. Notice in eq. \eqref{eq:SE8} how $m m \to m m$ does not exchange $m_4$ and how $M M \to M M$ does not exchange $m_3$.
\par
In summary:
\begin{itemize}
\item The $t$- and $u$-channel exchanges of $m_3$ cancel the $s$-channel exchange of $M$.
\item The $t$- and $u$-channel exchanges of $m_4$ cancel the $s$-channel exchange of $m$. 
\item There are no $s$-channel exchanges of $m_3$ and $m_4$, since $g_{M M m_3} = g_{m m m_4} = 0$.
\item The $t$- and $u$-channel exchanges of $M$ cancel the $t$- and $u$-channel exchanges of $m$ whose residue is anomalous according  to eq. (12) in the main text.
\end{itemize}

\begin{widetext}

\subsection*{Coleman-Thun double pole of $ MM \to M M$}

As seen in eq. \eqref{eq:SE8} the $MM \to MM$ S-matrix of the $E_8$ contains a double pole at $s = 4M^2 - M^4/m^2$. In units of $m = 1$ its residue reads

\begin{align}
&(s- a)^2\;T^{E_8}_{M M \to M M}(s \to a) = \notag \\
&-60 \sqrt{4031075805785+1802751904838 \sqrt{5}+10 \sqrt{30 \left(4844686384163194390099 \sqrt{5}+10833048084656563254865\right)}} 
\label{eq:E8dp}
\end{align}

Eq. (24) of the main text for the residue reads, in the notation here:
\begin{align}
&(s-a)^2\; T_{M M \to M M}(s \to a) =  { 2 \mathcal{N}_m ^2 g_{m m M}^4  \varrho(a) S_{m m \to m m}(a)} = \notag \\
&= 60 \sqrt{224644155305+100463920402 \sqrt{5}+10 \sqrt{30 \left(15045755001742455659 \sqrt{5}+33643330956703597625\right)}}
\label{eq:npdp}
\end{align}
which does not match with the $E_8$ double pole \eqref{eq:E8dp}. What is missing is the contribution from another overlapping singularity. Namely, from the box diagram in Fig. 3 of the main text. This diagram has anomalous thresholds in the $s-$ and $t-$ channels which as $M \to (1+\sqrt{5})/2$ overlap and create a douple pole at $s \to a$. Let us see this explicitly.

Let us first consider the anomalous threshold in the $s$-channel. Using the Cutkosky rules we can write down the unitarity equation for $M < \sqrt{2} m$ as
\be
\mathrm{Disc}_s T^{abox}_{M M \to M M}(s) = 2 \rho_{mm}(s) \left[- {\mathcal{N}_m g_{m m M}^2 \over s - a} \right] \left[ - {\mathcal{N}_M g_{m M M}^2 \over s - b} \right] \Theta(s - 4m^2)
\ee
where the terms in brackets correspond to the tree level exchanges of $m$ and $M$, as given by eqs. \eqref{eq:pole1} and \eqref{eq:simb}
Giving
\be
T^{abox}_{M M \to M M}(s) = 2 \mathcal{N}_m \mathcal{N}_M\, g_{m m M}^2 g_{m M M}^2 \left[{\varrho_{mm}(s) \over (s -a)(s-b)} - {\varrho_{mm}(a) \over (s -a)(a-b)} - {\varrho_{mm}(b) \over (b -a)(s-b)} \right] + \dots
\ee
where '$\dots$' include regular pieces and contributions with $t$-channel branch cuts.
\par
Performing the analytic continuation to $M > \sqrt{2} m$ we again find the middle piece changes $\varrho_{mm}(a) \to - \varrho_{mm}(a)$ and the $s \to a$ pole no longer cancels. So we get
\par
\be
\text{s-channel:} \qquad
T^{abox}_{M M \to M M}(s \to a) =  4 \mathcal{N}_m \mathcal{N}_M\, g_{m m M}^2 g_{m M M}^2 {\varrho_{mm}(a) \over (s - a)(a - b)}, \qquad M > \sqrt{2} m
\ee
Note that in the limit $M \to (1 + \sqrt{5})/2$, we have $b \to a$ and we find a divergence. Expanding $M = (1 + \sqrt{5})/2 + \epsilon$ in the above we find
\be
\text{s-channel:} \qquad
T^{abox}_{M M \to M M}(s \to a) =  \left[{1 \over \epsilon} \,{1 \over s - a} \, {2 \over \sqrt{5}}  - {8 \over \sqrt{5}} {1 \over (s - a)^2} \right] g_{mm M}^2 g_{m M M}^2 \, \varrho_{mm}(a)
\label{eq:sabox}
\ee
We will now see how this divergence gets canceled out from the $t$-channel contribution. 
\par
The $t$-channel unitarity cut gives
\be
\label{eq:unisolCat}
\mathrm{Disc}_t T^{abox}_{M M \to M M}(t) = 2 \rho_{mM}(t) \, \big[ T^{tree}_{M M \to M m}(t)\big]^2 \, \Theta(t - (m+M)^2)
\ee
where
\be
\label{eq:treeMMmM}
T^{tree}_{M M \to M m}(t) &\equiv - {g_{m m M} g_{m M M} \over \bar{s}(t) - m^2} - {g_{m m M} g_{m M M} \over \bar{u}(t) - m^2} \\
&= - {  g_{m m M} g_{m M M} P(t) \over (t- t_-)(t- t_+)}
\ee
with $P(t) =  (m^2 - 3M^2 + t) t /m^2$ and 
\be
\label{eq:tminus}
t = t_\pm = {3 M^2 \over 2} \pm {\sqrt{-4m^6 M^2 + 17 m^4 M^4 - 4m^2 M^6} \over 2 m^2}
\ee 
where we made use $d = 2$ kinematics for $M M \to m M$ scattering to find $\bar{s}(t)$ and $\bar{u}(t)$:
\be
\bar{s}(t), \; \bar{u}(t) = {1 \over 2} \left(m^2 + 3M^2 - t \pm {\sqrt{t(t-4M^2)} \sqrt{m^4+(M^2-t)^2 - 2m^2(M^2+t)} \over t} \right).
\ee
The motion of the singularity $4M^2 - t_+$ is plotted in Fig. \ref{fig:E8finetuning}. As $M \to {1 + \sqrt{5} \over 2} m$ it comes in contact with the $m M$ threshold and becomes an anomalous threshold for $M > {1 + \sqrt{5} \over 2} m$.\footnote{In fact, as $M \to {1 + \sqrt{5} \over 2}m$ we also have $t_- \to m^2$ which gets canceled in the $E_8$ model by an $s$-channel exchange of $m^2$ of $MM \to m M$.} 
\par
We now plug \eqref{eq:treeMMmM} back into \eqref{eq:unisolCat}. Proceeding as before, we must cancel the anomalous threshold for $M < \sqrt{2}m$, so we must have that
\be
\label{eq:aboxt}
T^{abox}_{M M \to M M}(t) = {2 g_{m m M}^2 g_{m M M}^2 \over (t - t_+)^2}  \left[ { \varrho_{mM}(t) P^2(t) \over (t- t_-)^2} - { \varrho_{mM}(t_+) P^2(t_+) \over (t_+ - t_-)^2} - (t - t_+) \, {d \over d t} \bigg( 
{ \varrho_{mM}(t) P^2(t) \over (t- t_-)^2} \bigg)\!_{t \to t_+} \right] + \dots
\ee
where '\dots' are regular and pieces coming from the $s$-channel.
\par
In $d=2$ we have $t = 4M^2 - s$. In the limit $M \to {1  + \sqrt{5} \over 2}$ the normal threshold at $t(s) = (m + M)^2$  will overlap with the anomalous threshold $s \to a$ from the $s$-channel (see fig. \ref{fig:E8finetuning}):
\be
t\text{-channel } mM \text{ threshold}: \qquad &t(s) = (m + M)^2 \implies \notag \\
&\implies s = 4M^2 - (m + M)^2 \to a, \;  \text{ if } \; M \to {1 + \sqrt{5} \over 2} m.
\ee
Therefore the phase space factor $\varrho_{m M}(t)$ will diverge. We see that 
\be
\varrho_{m M}(t_+) = -\frac{\sqrt{1-\frac{2}{\sqrt{5}}}}{4 |\epsilon| } + O(\epsilon^0), \qquad M = {1 + \sqrt{5} \over 2} + \epsilon.
\ee
Carefully expanding $M \to {1 + \sqrt{5} \over 2} + \epsilon$ we find the following contribution from the $t$-channel to the singularity at $s \to a$ of the box diagram:
\be
\text{t-channel:} \qquad T^{abox}_{MM \to MM}(s \to a) &= \left[-{1 \over \epsilon} \,{1 \over s - a} \,{4 \sqrt{2} \sqrt{5+\sqrt{5}} \over 10} \,  - {\sqrt{130+38 \sqrt{5}}\over 5 \, (s - a)^2} \right] g_{mm M}^2 g_{m M M}^2.
\ee
Putting this together with the $s$-channel piece \eqref{eq:sabox} we find that the $1/\epsilon$ pieces cancel out and we get an additional contribution to the double pole residue given by eq. (25) in the main text. Plugging in the cubic couplings from the $E_8$ model given by eqs. \eqref{eq:gmmM} and \eqref{eq:gmMM}, we find
\begin{align}
&(s-a)^2\; T^{abox}_{M M \to M M}(s \to a) =  -  {8 \sqrt{5}\,\varrho_{mm}(a) + \sqrt{130+38 \sqrt{5}}\over 5} {g_{mmM}^2 g_{mMM}^2 \over (s - a)^2}. = \notag \\
&= -60 \sqrt{6158935786330+2754359817458 \sqrt{5}+40 \sqrt{30 \left(706830217586335483049 \sqrt{5}+1580520415074013466285\right)}}.
\label{eq:E8abox}
\end{align}
The reader may now check that this piece together with \eqref{eq:npdp} reproduces the $E_8$ model double pole residue given by eq. \eqref{eq:E8dp}.

\section{Kinematics, unitarity and Legendre functions}
\label{app:kinematics}

Here we collect several technical details regarding kinematics, unitarity and Legendre functions.

\subsection*{Scattering angle}
Let us relate $t$ and $u$ with the cosine of the scattering angle in the center of mass frame. We let $\vec{p}$ and $E_p$ be the momentum and energy of a particle of mass $M$ and $\vec{k}$ and $E_k$ be the momentum and energy of a particle of mass $m$. Then, the cosine of the scattering angle is 
\be
z \equiv {\vec{p} \cdot \vec{k} \over |\vec{p}| |\vec{k}|}.
\ee
For the scattering $m m \to M M$, in the center of mass we have $E_p = E_k = \sqrt{s}/2$. Further using $|\vec{p}| = \sqrt{E_p^2 - M^2}$ and $|\vec{k}| = \sqrt{E_k^2 - m^2}$ we get
\be
\label{eq:tz}
t(z) &= (p - k)^2 = M^2 + m^2 - 2 E_p E_k  + 2  |\vec{p}| |\vec{k}| z \notag \\
&= m^2 + M^2 + {-s + z \sqrt{s - 4m^2} \sqrt{s - 4M^2} \over 2},
\ee
where, also, $u(z) = t(-z)$.
\par
Now, in $d=2$ there is only forward or backward scattering, i.e. $z = \pm 1$, so $t = t(s)$ and $u = u(s)$ are fixed in terms of $s$,
\be
\label{eq:ts2d}
t(s) &= m^2 + M^2 + {-s + \sqrt{s - 4m^2} \sqrt{s - 4M^2} \over 2}, \\
u(s) &= m^2 + M^2 + {-s - \sqrt{s - 4m^2} \sqrt{s - 4M^2} \over 2} \label{eq:us2d}.
\ee

\par
Relation \eqref{eq:tz} inverts to
\be
\label{eq:zt}
z(t)= \frac{s -2 m^2-2 M^2+ 2 t}{\sqrt{s-4 m^2} \sqrt{s-4 M^2}}.
\ee
\par
\subsection*{Extended unitarity for the amplitudes}

In terms of the amplitudes, unitarity across the $2m$ cut reads [61], for the $m m \to M M$ process,
\be
\label{eq:elauniB}
&\mathrm{Disc}_s B(s,t) = {1 \over 8 (4 \pi)^2} \sqrt{s - 4m^2 \over s} \notag \\
&\times  \iint_{-1}^1 dz' dz'' \, \mathcal{P}(z,z',z'') A(s,t(z')) B^*(s,t(z''))
\ee
where $z$ and $z''$, scattering angles of $m m \to MM$, are related to $t(z)$ and $t(z'')$ via \eqref{eq:zt} and $z' = 1 + {2 t(z') \over s-4m^2}$ is the usual relation for the scattering angle of $m m \to m m$.
\par
Likewise, for the $MM \to MM$ process,
\be
\label{eq:elauniC}
&\mathrm{Disc}_s C(s,t) = {1 \over 8 (4 \pi)^2} \sqrt{s - 4m^2 \over s} \notag \\
&\times  \iint_{-1}^1 dz' dz'' \, \mathcal{P}(z,z',z'') B(s,t(z')) B^*(s,t(z'')),
\ee
where $z = 1 + {2 t(z)  \over s-4M^2}$ is the usual relation for the scattering angle of $MM \to M M$ and $z'$ and $z''$, scattering angles of $m m \to M M$ are related to $t(z')$ and $t(z'')$ via eq. \eqref{eq:zt}.
\par
The 2-particle kernel reads (following conventions of [61])
\be
\label{eq:P}
\mathcal{P}(z,z',z'') = {2 \, \Theta(1 - z^2 - z'^2 - z''^2 +2 z z' z'') \over \sqrt{1 - z^2 - z'^2 - z''^2 +2 z z' z''}}
\ee
Eqs. \eqref{eq:elauniB} and \eqref{eq:elauniC}, because of the theta function in \eqref{eq:P} only hold for real $t$ in the scattering angle region $-1 < z < 1$. These equations can however be continued in $t$ [2,3] and be expressed in manifestly analytic form. Say for $MM \to MM$,
\be
\label{eq:elauniC2}
&\mathrm{Disc}_s C(s,t) = {1 \over 8 (4 \pi)^2} \sqrt{s - 4m^2 \over s} \left({1 \over 2 \pi i} \right)^2\notag \\
&\times \oint_{-1}^1 dz' \oint_{-1}^1 dz'' K(z,z',z'') B(s,t(z')) B^*(s,t(z'')),
\ee
with Mandelstam kernel $K(z,z',z'')$ which has discontinuity [61]
\be
\label{eq:DK}
\mathrm{Disc}_z K(z,z',z'') = 4 \pi^2 {\Theta(z - z_+(z',z'')) \over \sqrt{(z - z_+(z',z''))(z - z_-(z',z''))}}
\ee
with 
\be
z_\pm(z',z'') = z' z'' \pm \sqrt{(z'^2 -1)(z''^2 -1)}.
\ee
Equation \eqref{eq:DK} can be integrated to give
\be
\label{eq:K}
K(z,z',z'') = {8 \pi \over \sqrt{(z_+ - z)(z - z_-)}} \,\mathrm{arctan} \sqrt{z - z_-\over z_+ - z}.
\ee
Note that $K(z,z',z'')$, as a function of $z$, has the same analyticity structure as the `bubble' diagram $W(s)$, in eq. \eqref{eq:W}, as a function of $s$.

\subsection*{Partial waves and Legendre functions}

The partial wave decomposition of the amplitude $B(s,t)$ reads [61]
\be
\label{eq:pwd}
B(s,t) = 16 \pi \sum_{J= 0}^\infty (1 + 2 J) \, P_J(z(t)) \, B_J(s)
\ee
where $P_J(z)$ is the Legendre polynomial and $z(t)$ the cosine of the scattering angle given by \eqref{eq:zt}. Analogous expressions hold for $A(s,t)$ and $C(s,t)$.
\par
Now, eq. \eqref{eq:pwd} can be inverted for the partial wave in the usual way. Alternatively, we can exploit analyticity in $t$ by considering the Legendre function of the second kind $Q_J(z)$. This turns eq. (14) in the main text into the so-called Froissart-Gribov representation [61]:
\be
B_J(s) = {1 \over 16 \pi} \oint_{[-1,1]} {dz \over 2\pi i} \,Q_J(z) \, B(s,t(z)) 
\ee
where the contour is counter-clockwise. Using eq. \eqref{eq:zt} we have instead
\be
\label{eq:FG}
B_J(s) =& {1 \over 8 \pi \sqrt{s-4 m^2} \sqrt{s-4 M^2}} \notag \\
&\;\;\times \oint {dt \over 2\pi i} \,Q_J(z(t)) \, B(s,t) 
\ee
Now, the partial wave decomposition diagonalizes unitarity. This is made explicit by an interesting identity between the 2-particle kernels $\mathcal{P}$ and $K$ and the Legendre functions. In particular, for the latter [61]:
\be
\label{eq:PQQ}
K(z,z',z'') = 4 \pi \sum_{J=0}^\infty (2J + 1) \,P_J(z) \, Q_J(z') \, Q_J(z'')
\ee
Plugging \eqref{eq:PQQ} into \eqref{eq:elauniC2} leads to partial wave unitarity and likewise for $A(s,t)$ and $B(s,t)$.

\end{widetext}

\bibliographystyle{JHEP}
\bibliography{papers} 

\end{document}